\documentclass[PAPER, cernpreprint, atlasdraft=false, texlive=2023, USenglish, texmf, orcidlogo]{atlasdoc}
 
\usepackage{placeins}
\usepackage{amsmath}
\usepackage{atlaspackage}
\usepackage{atlasbiblatex}
 
\usepackage{atlasphysics}
\usepackage{atlasheavyion}
\usepackage{atlasjetetmiss}

\graphicspath{{logos/}{figures/}}
 
\usepackage{ANA-HION-2019-09-PAPER-defs}
 
\newcommand{\lR}{large-radius}

\newcommand{\dRonetwo}{\ensuremath{\Delta R_{12}}\xspace}

\newcommand{\pythiaeight}{\textsc{Pythia}8}
\newcommand{\ETfcal}{\mbox{$\Sigma E_{\mathrm{T}}^{\mathrm{FCal}}$}}

\AtlasTitle{Measurement of suppression of large-radius jets and its
dependence on substructure in Pb+Pb collisions at $\sqrt{s_\mathrm{NN}} = 5.02$~TeV with the ATLAS detector}

\AtlasAbstract{
This letter presents a measurement of the nuclear modification factor of large-radius jets in $\sqrt{s_\mathrm{NN}} = 5.02$~TeV Pb+Pb collisions by the ATLAS experiment. The measurement is performed using 1.72 nb$^{-1}$ and 257~pb$^{-1}$ of Pb+Pb and $pp$ data, respectively. The large-radius jets are reconstructed with the anti-$k_{t}$ algorithm using a radius parameter of $R = 1.0$, by re-clustering anti-$k_{t}$ $R = 0.2$ jets,
and are measured over the transverse momentum  ($p_{\mathrm{T}}$) kinematic range of $158 < p_{\mathrm{T}} < 1000$~GeV and absolute pseudorapidity $|y|<2.0$.
The large-radius jet constituents are further re-clustered using the $k_{t}$ algorithm in order to obtain the splitting
parameters, $\sqrt{d_{12}}$ and $\Delta R_{12}$, which characterize the transverse momentum scale and angular separation for the hardest splitting in the jet, respectively. The nuclear modification factor, $R_{\mathrm{AA}}$, obtained by comparing the Pb+Pb jet yields to those in $pp$ collisions, is measured as a function of jet transverse momentum ($p_{\mathrm{T}}$) and $\sqrt{d_{12}}$ or $\Delta R_{12}$. A significant difference in the quenching of large-radius jets having single sub-jet and those with more complex substructure is observed. Systematic comparison of jet suppression in terms of $R_{\mathrm{AA}}$ for different jet definitions is also provided. Presented results support the hypothesis that jets with hard internal splittings lose more energy through quenching and provide a new perspective for understanding the role of jet structure in jet suppression.
}
 
\author{The ATLAS Collaboration}
 
\AtlasJournalRef{\PRL 131 (2023) 172301}
\AtlasDOI{10.1103/PhysRevLett.131.172301}
 
\PreprintIdNumber{CERN-EP-2022-250}

\hypersetup{pdftitle={ATLAS document},pdfauthor={The ATLAS Collaboration}}
 
\begin{document}
 
\maketitle
 
Heavy-ion collisions at high energies lead to the creation of matter
composed of unconfined quarks and gluons
known as quark-gluon  plasma (QGP). Studies of the collective
expansion of the QGP~\cite{Heinz:2013th} have demonstrated that the plasma is
strongly coupled. A major goal of the experimental high-energy nuclear
physics programs at the Relativistic Heavy Ion Collider (RHIC) and Large Hadron Collider (LHC) is to understand how the strong
coupling of the QGP arises from a theory that is asymptotically
free~\cite{Busza:2018rrf,Gelis:2021zmx}.
Jets, collimated sprays of hadrons originating in hard scattering
processes, represent a key tool for studying the microscopic interactions
between color charges within the QGP that ultimately determine its properties.
Those interactions can alter the energy and angular
distributions of constituents within the jets --- a phenomenon referred
to as ``jet quenching''~\cite{Qin:2015srf,Blaizot:2015lma}.
As a result, the jet yields are suppressed and jet properties are significantly modified in heavy-ion collisions relative to those measured in elementary proton--proton (\pp) collisions.

At the LHC, jet quenching was studied using many observables~\cite{Cunqueiro:2021wls}. Jet substructure observables~\cite{Marzani:2019hun}, constructed
from measured jet constituents, are versatile tools to measure the
changes in jet properties due to jet
quenching. These observables, calculated separately for each jet using
a variety of algorithms, were originally motivated by studies of
highly Lorentz-boosted massive resonances in elementary
collisions~\cite{Butterworth:2008iy}.
Jet grooming algorithms that remove soft and wide-angle radiation can help mitigate the impact of backgrounds from multiple collisions within one bunch crossing (pileup) or the underlying event (UE) on these observables.
They also can be used to separate hard
components of a parton shower, for example, sub-jets, in a jet with multi-prong
structure~\cite{Dasgupta:2013ihk,Larkoski:2014wba}. In the context of jet quenching, these methods enable distinguishing sub-jets resulting from hard splittings in the
parton shower from soft medium-induced
radiation~\cite{Blaizot:2012fh}, soft particles resulting from
jet-induced medium excitations~\cite{Tachibana:2019hrn}, and the
UE. Furthermore, the dependence of the jet quenching on such
splittings, and more generally, the complexity of the parton
shower~\cite{Andrews:2018jcm} can be studied
using substructure techniques.
 
One specific phenomenon connected with the parton shower complexity is color (de)coherence~\cite{Mehtar-Tani:2010ebp,MehtarTani:2011tz,CasalderreySolana:2012ef}. In the vacuum, the quantum interference in the parton cascade suppresses wide-angle gluon emissions~\cite{CDF:1994zkl,CMS-SMP-12-010,lhcbcollaboration2022multidifferential}. This phenomenon in the medium-induced parton cascade is believed to largely dictate the magnitude of the energy loss~\cite{Mehtar-Tani:2017web,Caucal:2018dla}.
In-medium jet evolution is then characterized by a vacuum-like parton cascade whose constituents are either resolved by the medium due to color decoherence leading to a radiation of individual color charges \cite{Baier:1996kr,Gyulassy:2000er,Chien:2015hda} or remain unresolved and radiate coherently as a single color charge.
This induces a dependence of the observed jet suppression on the structure of splittings, which may be experimentally accessed and tested by substructure techniques~\cite{Andrews:2018jcm}.
 
Jet substructure was measured in lead-lead (\PbPb) collisions at the LHC using several observables: the momentum ratio of two leading sub-jets, $z_g$~\cite{CMS-HIN-16-006,ALICE:2019ykw,ALICE:2021obz},
the groomed jet radius, $\theta_g$~\cite{ALICE:2021obz}, the groomed jet mass~\cite{CMS-HIN-16-024}, the number of branches obtained in the iterative declustering of the jet~\cite{ALICE:2019ykw}, the $N$-subjetiness~\cite{ALICE:2021vrw}, and momentum fraction of sub-jets \cite{ALICE:2022vsz}. Typically the per-jet normalized distributions of substructure observables are measured in \PbPb\ collisions and compared with the same quantity measured in \pp\ collisions or to that obtained from Monte Carlo (MC) simulations. A recent measurement of $z_g$, fully corrected for detector effects and with
a robust grooming method to reduce the impact of fluctuating backgrounds on the sub-jet selection, reported no significant change of the $z_g$ distribution in \PbPb\ collisions~\cite{ALICE:2021obz}.  A narrowing of the $\theta_g$ distribution was observed in \PbPb\ collisions~\cite{ALICE:2021obz} which may be due to the color coherence effects or due to the
difference in the relative suppression of quark- and
gluon-initiated jets~\cite{Ringer:2019rfk}.  No change of the groomed mass distribution for the core of the jet was observed, and only a
hint of an increase for jets with large jet mass was
seen~\cite{CMS-HIN-16-024}.  Furthermore, no significant change in the per-jet yields  of the two-pronged structure was observed in \PbPb\ collisions
relative to the MC-based reference in the measurement of
$N$-subjetiness~\cite{ALICE:2021vrw}.
 
One important conceptual issue associated with jet substructure measurements is that the QGP may not directly modify the hard splittings of a jet since their formation time may be much shorter than the formation time of the QGP~\cite{Caucal:2018dla}.
Thus, modifications of substructure distributions due to quenching may depend on the structure of the part of the parton shower that is dictated by the vacuum.
Rather than measuring the per-jet normalized distributions of a substructure variable, which are resulting from an admixture of vacuum-driven and medium-driven substructure, it may be advantageous to, instead, quantify the suppression of jets having different substructure using the jet nuclear modification factor, \RAA~\cite{HION-2013-06,CMS-HIN-13-005,Acharya:2020,STAR:2020xiv}.
 
This Letter provides the first observation and quantification of the dependence of the \lR\ jet suppression on the jet substructure. The \lR\ ($R=1.0$) jets are reconstructed using a new method of re-clustering from small-radius ($R=0.2$) jets.
The measurement is done with the ATLAS detector differentially in jet transverse momentum (\pt) and also in two specific substructure observables.
The jet suppression is quantified in terms of jet \RAA. The two substructure observables are the splitting parameter, $\sqrt{d_{12}}$, and the angular separation, \dRonetwo, which characterize the transverse momentum scale and the angular separation, respectively, for the hardest splitting in the jet. 
The differential jet suppression measurement allows direct quantification of
the difference between the energy loss of jets with single sub-jet and the energy loss of jets with more complex structure involving early hard splittings. This quantification was not directly accessible in previous measurements. The use of \lR\ jets then delimits the measured kinematics of the internal splitting. The results also provide a direct quantification of cone-size dependence of energy loss. Detailed quantification of jet suppression as a function of angular separation at small opening angles using tracks associated with calorimeter small-radius jets follows in a separate study by ATLAS~\cite{HION-2021-09}.
 
The principal components of the ATLAS detector~\cite{PERF-2007-01} used in this measurement are the inner tracking detector, electromagnetic and hadronic calorimeters, and the online trigger system. The inner tracking detector is surrounded by a thin superconducting solenoid providing a \SI{2}{\tesla}
axial magnetic field and it covers the pseudorapidity range \(|\eta| < 2.5\).\footnote{ATLAS uses a right-handed coordinate system with its origin at the nominal interaction point (IP) in the center of the detector and the $z$-axis along the beam pipe. The $x$-axis points from the IP to the center of the LHC ring, and the $y$-axis points upward. Cylindrical coordinates $(r,\phi)$ are used in the transverse plane, $\phi$ being the azimuthal angle around the $z$-axis. The pseudorapidity is defined in terms of the polar angle $\theta$ as $\eta=-\ln\tan(\theta/2)$. The rapidity is defined as \(y\) = \(0.5 \ln[(E+p_{z})/(E-p_{z})]\) where \(E\) and \(p_{z}\) are the energy and \(z\)-component of the momentum along the beam direction, respectively.} It consists of silicon pixel, silicon microstrip, and transition radiation tracking detectors. Lead/liquid-argon (LAr) sampling calorimeters provide electromagnetic (EM) energy measurements with high granularity. A steel/scintillator-tile hadron calorimeter covers the central pseudorapidity range (\(|\eta| < 1.7\)).
Liquid-argon calorimeters with separate EM and hadronic compartments instrument the endcap (up to $|\eta| = 3.2$) and forward (FCal, up to $|\eta| = 4.9$) regions.
Both the inner detector and calorimeter systems have a 2$\pi$ coverage in azimuth.
Two zero-degree calorimeters (ZDCs) are composed of four longitudinal layers of tungsten absorbers and quartz rods. They are situated in the far forward region $|\eta| > 8.3$ and primarily measure the spectator neutrons from the struck Pb nuclei.
An extensive software suite~\cite{ATL-SOFT-PUB-2021-001} is used in data simulation, in the reconstruction
and analysis of real and simulated data, in detector operations, and in the trigger and data acquisition systems of the experiment.

The analysis uses 1.72 nb$^{-1}$ of \PbPb\ data at $\sqn=5.02$~\TeV\  recorded in 2018 and 257 pb$^{-1}$ of \pp\ data collected in 2017 at the same center-of-mass energy. Events were selected using a combination of calorimeter-based jet triggers~\cite{ATL-DAQ-PUB-2019-001, TRIG-2016-01}. Both the \pp\ and \PbPb\ events are required to contain at least one primary vertex. All jets in the analysis are in a kinematic range where the jet trigger was fully efficient.
The \PbPb data contain only a small fraction of events (<0.5\%) with multiple collisions per bunch crossing which are further suppressed using the anti-correlation of signal from the ZDC and FCal. The \pp data were collected with typically $1.4-4.4$ inelastic interactions per bunch crossing. No pileup rejection is applied in the analysis of \pp data.
 
The centrality of \PbPb events is defined using the total transverse energy measured in the FCal, $\ETfcal$~\cite{HION-2016-07,HION-2018-06}. Events in \PbPb data are classified into four centrality intervals, ordered from the most central to the most peripheral: $0-10\%$, $10-30\%$, $30-50\%$, and $50-80\%$. The values of the mean nuclear thickness function with their uncertainties, $\langle\TAA\rangle$ (used as an input to \RAA), are evaluated in each centrality interval by a Glauber model analysis of the \ETfcal\ distribution~\cite{Miller:2007ri,Loizides:2014vua}.

Several MC simulation samples are used to evaluate the performance of the analysis procedure and to correct the measured distributions for detector effects. The \pp\ MC sample uses $4\times10^7$ \pythiaeight~\cite{Sjostrand:2014zea} jet events at $\sqrt{s} =5.02$~TeV with the A14 ATLAS tune~\cite{ATL-PHYS-PUB-2014-021} and the NNPDF23LO parton distribution functions (PDF)~\cite{Ball:2012cx}. Pileup from additional \pp\ collisions is generated by \pythiaeight, with parameter values set to the A2 tune~\cite{ATL-PHYS-PUB-2011-014} and using the MSTW2008~\cite{Martin:2009iq} PDF set, with a distribution of the number of extra collisions matching that of data.
 
The \PbPb\ MC sample uses $4\times10^7$ \pp\ \pythiaeight\ events with the same tune and PDFs as in \pp\ MC samples that are overlaid on top of events from a dedicated sample of \PbPb\ data events. This sample was recorded with a combination of minimum-bias and total energy triggers requiring 1.5 TeV or 6.5 TeV to
enhance the number of central collisions. This ``MC overlay'' sample was re-weighted on an event-by-event basis such that it has the same centrality distribution as the jet-triggered data sample. The detector response in all MC samples was simulated
using \GEANT~\cite{Agostinelli:2002hh,SOFT-2010-01}.

First, jets with radius parameter $R=0.2$ are reconstructed using the anti-\kt algorithm~\cite{Cacciari:2008gp,Fastjet} from calorimeter energy deposits as described in Ref.~\cite{HION-2017-10}. The jet kinematics are corrected event-by-event for the contribution from the UE particles, and are calibrated using simulations of the calorimeter response and in situ measurement of the absolute energy scale~\cite{PERF-2016-04}.
 
The \lR\ jets are defined by clustering the small-radius, $R = 0.2$, jets with $\pt > 35$~GeV and $|\eta|<3.0$ using anti-$k_t$ algorithm with radius parameter $R=1.0$~\cite{Nachman_2015}. These requirements and procedure limits the impact of UE on the measurement but prohibits recovering the quenched jet energy~\cite{Blaizot:2012fh} transferred outside of the $R = 0.2$ sub-jets. This procedure allows to perform the measurement in a wide kinematic range and reduces systematics uncertainties. The $k_t$ jet finding algorithm~\cite{Fastjet,Catani:1993hr} is used to re-cluster constituents of $R = 1.0$ jet (i.e. $R=0.2$ jets) to obtain two observables of interest, \dRonetwo and \Donetwo. These are defined as the
angular separation and the splitting parameter of the $k_t$ algorithm
calculated for two jets before the final clustering step of $R=1.0$ jet
\begin{equation}
\nonumber
\dRonetwo  = \sqrt{ \Delta y_{12}^2 + \Delta \phi_{12}^2},~
\Donetwo = \mathrm{min}({\pt}_{1}, {\pt}_{2}) \times \dRonetwo.
\end{equation}
 
The \lR\ jet yields in \PbPb\ collisions and jet cross-section in \pp\
collisions measured in the kinematic range of $158 < \pt < 1000$~GeV and $|y|<2.0$ are evaluated
inclusively and differentially in each of the two substructure observables.   The \kt\ clustering of \lR\ jet constituents with only a single sub-jet
defaults to $\Donetwo = 0$ and $\dRonetwo =0$. These jets, which consist of a single $R=0.2$ jet, centered on the $R=1.0$ jet axis with no other clustered activity in the \lR\ cone, populate only the first interval in \Donetwo\ and \dRonetwo. The fraction of reconstructed \lR\ jets with a single sub-jet in \pp\ collisions is 75\% in the \pt\ interval of $158-200$~\GeV\ and 62\% in the \pt\ interval of $316-500$~\GeV. More details of the sub-jet multiplicity can be found in the Appendix. The jet performance of \lR\ re-clustered jets was analyzed using MC samples. The jet energy scale (JES) was found to be within a 3\% range from 1.

The measured inclusive and differential distributions are corrected for detector effects by the iterative Bayesian unfolding
method~\cite{DAgostini:1994zf,Adye:2011gm} in one dimension and
two dimensions, respectively, to return the distributions to the particle level. The unfolding accounts for the effects of bin migrations due to the jet energy resolution (JER) and JES. It also corrects the combinatoric sub-jet contribution originating from fluctuations of the UE and from jets from
different hard partonic interactions in the same \PbPb\ collision resulting in migration in the substructure observable. To better represent the data, the simulated distributions are re-weighted along the generator-level jet \pT, \Donetwo,  and \dRonetwo\ axes by the reconstruction-level data-to-simulation ratio before the unfolding. The number of iterations in the unfolding was chosen
such that the result is stable when changing the number of iterations while minimizing the amplification
of statistical uncertainties. Four iterations were used for inclusive $R=0.2$ and $R=1.0$ jet yields and cross-sections, while six and eight iterations were used in the unfolding of \Donetwo\ and \dRonetwo\ distributions, respectively.  Generator-level jets that do not match to a reconstructed jet passing the selection criteria are incorporated as an inefficiency correction after the unfolding.

A dominant source of systematic uncertainty is the uncertainty in the JES. For more central collisions, the uncertainty in JER is equally important. The systematic uncertainty
on the JES contains components from calorimeter response uncertainties
derived from in situ studies~\cite{PERF-2016-04}, a component connected with the JES in the \PbPb\ environment~\cite{ATLAS-CONF-2015-016}, components accounting for inaccuracies in the description of the relative abundances of jets initiated by quarks and gluons and their calorimetric response, and a component connected with the jet radius. The magnitude of the uncertainty in the \RAA\ from the JES uncertainty varies from 5\% to 20\% as a function of \pT, \Donetwo, \dRonetwo, and centrality. The primary component of the JER uncertainty is derived using an in situ technique involving studies of dijet energy balance~\cite{PERF-2011-04,ATLAS-CONF-2015-017}. The resulting uncertainty in \RAA\ reaches $\sim$20\% in 0--10\% central collisions at low \Donetwo and \dRonetwo, but it is typically below 5\%. The systematic uncertainty of the unfolding procedure is estimated by repeating the analysis with response matrices without the re-weighting to match the shapes of measured distributions in data and it is typically bellow 5\%. Other systematic uncertainties consist of the deviation between the unfolded result and the underlying generator-level distribution in simulation, the uncertainty in the determination of $\langle\TAA\rangle$ values~\cite{HION-2017-10}, and the uncertainty in the determination of the \pp\ luminosity~\cite{ATLAS-CONF-2020-023}.

\begin{figure}
\centering
\includegraphics[width=0.45\textwidth]{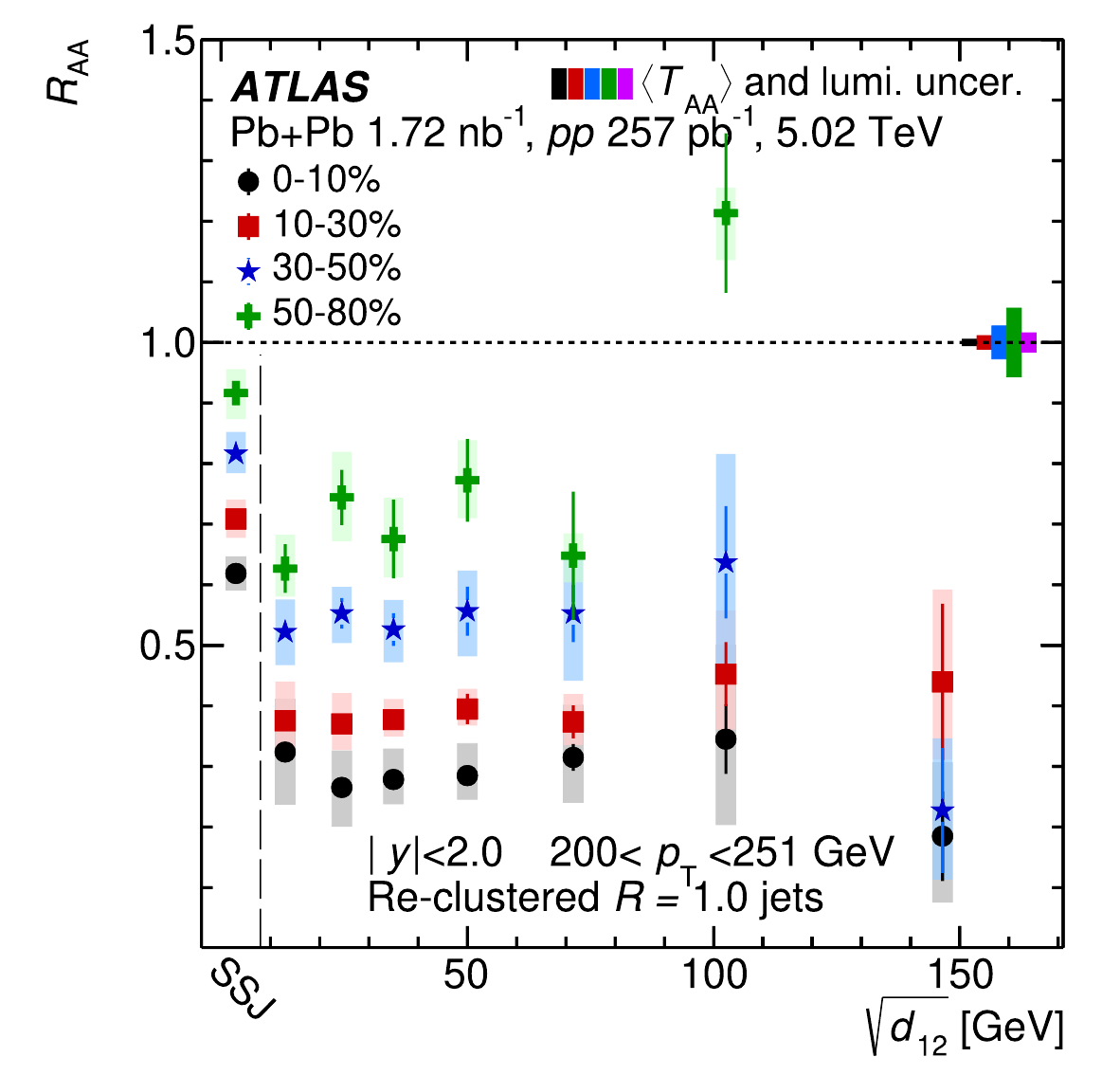}
\includegraphics[width=0.45\textwidth]{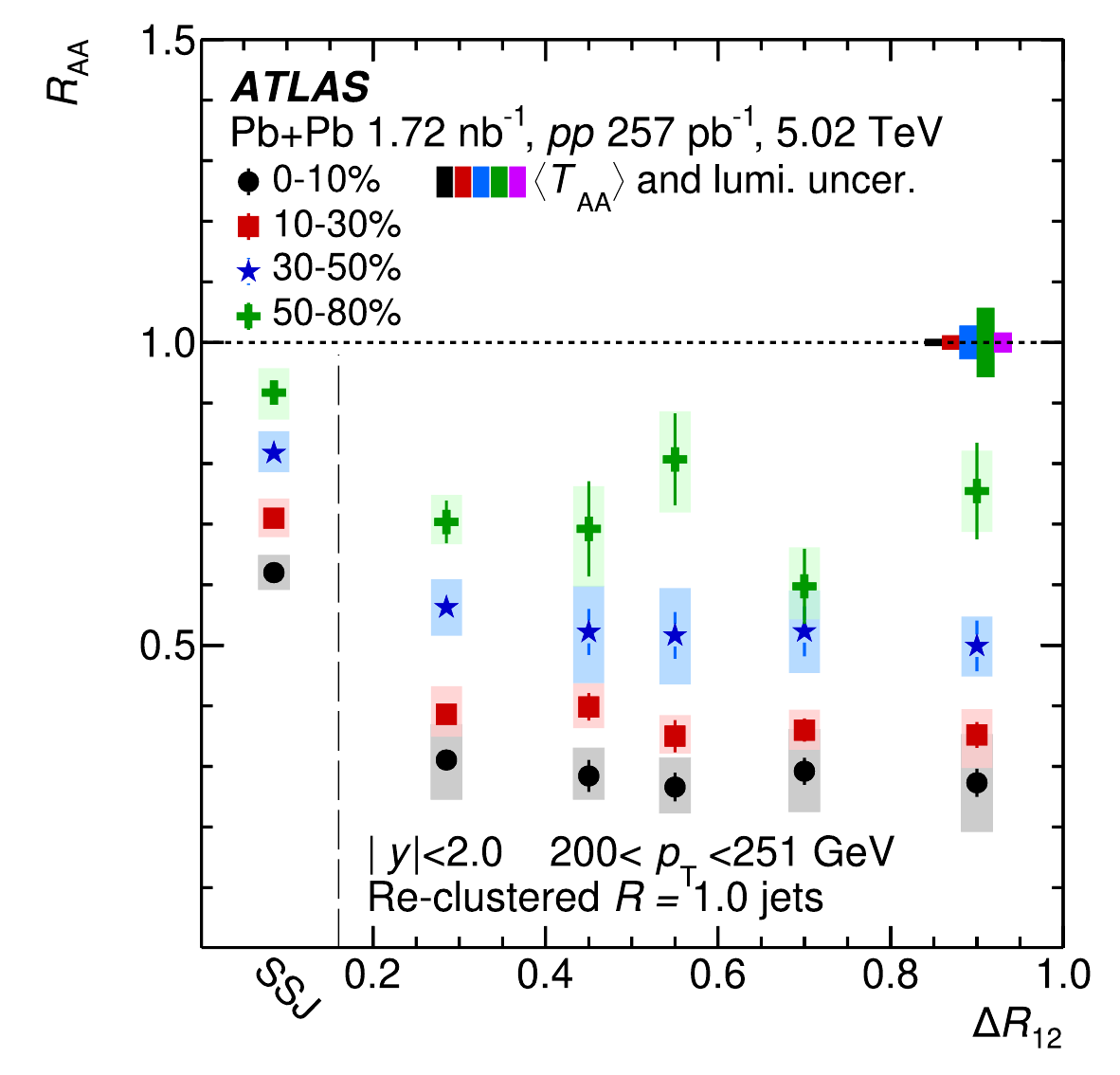}
\caption{
The values of \RAA for $R = 1.0$ re-clustered jets as function of \Donetwo (left) and \dRonetwo (right)  in four centrality intervals. The label `SSJ' on the $x$-axis identifies the single sub-jet configuration. The vertical bars on the data points indicate statistical uncertainties, the shaded boxes indicate systematic uncertainties. The fully correlated fractional uncertainties due to the luminosity and $\langle\TAA\rangle$ are represented by boxes at $\RAA=1$. One data point with a relative statistical uncertainty above 50\% is not displayed.
}
\label{fig:thetwo}
\end{figure}

The nuclear modification factor for \lR\ jets evaluated in the \pt\ interval of $200-251$~\GeV\ as a function of the \Donetwo\ is shown in the left panel of Figure~\ref{fig:thetwo}. The \RAA\ values for \lR\ jets with a single sub-jet are significantly larger compared with the \RAA\ for \lR\ jets with a more complex substructure having a non-zero \Donetwo. This observation is qualitatively consistent with the scenario in which the medium cannot resolve partonic fragments below a certain transverse scale~\cite{Casalderrey-Solana:2017mjg}. This result is also consistent with the previous measurement of correlated production of pairs of nearby jets in \PbPb\ collisions~\cite{HION-2013-10} where a larger suppression of neighboring jets compared with inclusive jets was observed.
For $\Donetwo>0$, the \RAA\ values are constant as a function of \Donetwo\ within uncertainties for all the centrality intervals.
 
The right panel of Figure~\ref{fig:thetwo} shows \RAA\ of \lR\ jets evaluated as a function of \dRonetwo for $\pt = 200-251$~\GeV. The trends seen for $\RAA(\dRonetwo)$ are the same as those seen for $\RAA(\Donetwo)$: suppression is significantly smaller for the single sub-jets case, and then constant within uncertainties for non-zero \dRonetwo\ values. The other \pt\ intervals spanning the range of $158 < \pT < 500$~\GeV\  show the same trends. These observations are in contrast with non-monotonic \pt\ and radial dependence of production of charged particles associated with jets measured using fragmentation functions or jet shapes~\cite{HION-2018-03,HION-2017-04,CMS-HIN-12-002}.
The measurement of the \kt\ splitting scale by \Donetwo\ and radial dependence of sub-jet suppression by \dRonetwo\ thus provide new information relative to previously measured transverse and radial structure of $R=0.4$ jets using charged particles.
The measurement also provides new information relative to the recent substructure measurement of $\theta_g$ distributions of $R=0.4$ jets \cite{ALICE:2021obz}. In that measurement, a difference in the slope of per-jet normalized $\theta_g$ distribution was seen in Pb+Pb with respect to $pp$ collisions, while here we report a difference in the suppression connected with particular sub-jet topology.
 
\begin{figure}
\centering
\includegraphics[width=0.75\textwidth]{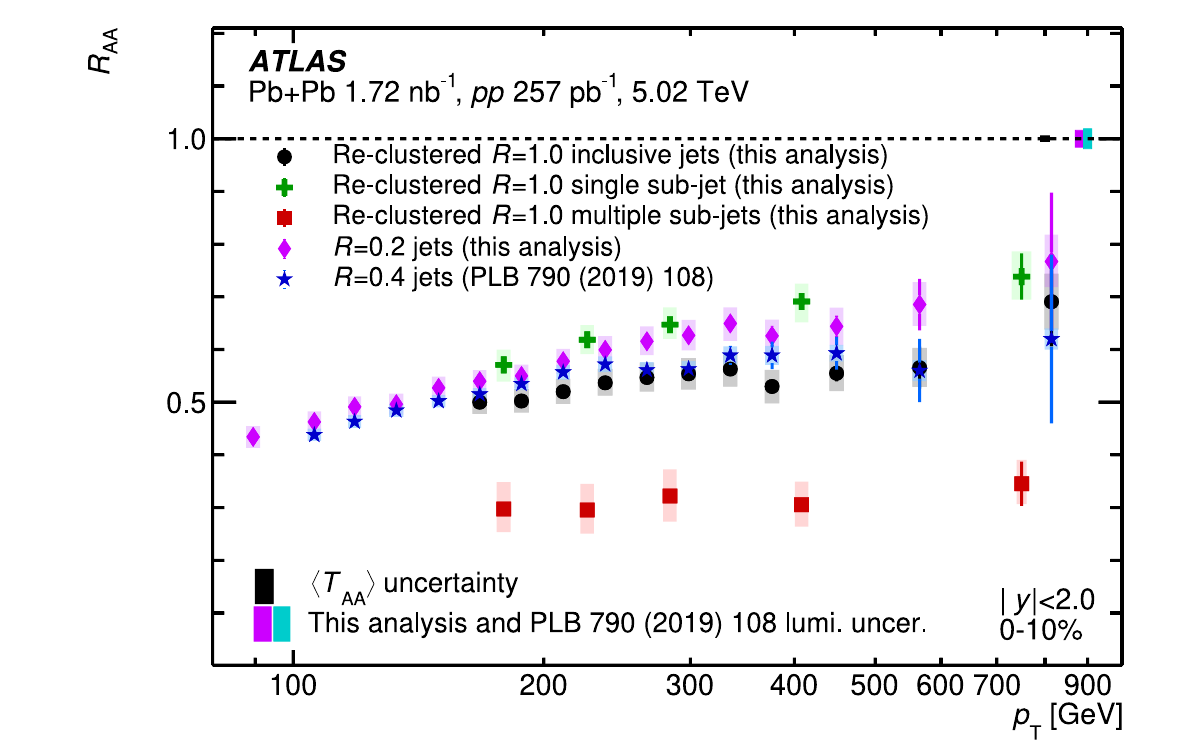}
\caption{Comparison of \RAA distributions evaluated in $0-10\%$ central collisions as a function of \pt\ for several jet definitions: $R=1.0$ re-clustered inclusive jets (circles), $R=1.0$ re-clustered jets with a single sub-jet (crosses), $R=1.0$ re-clustered jets with multiple sub-jets (squares), $R=0.2$ jets (diamonds), $R=0.4$ jets (stars)~\cite{HION-2017-10}.
The vertical bars on the data points indicate statistical uncertainties, the shaded boxes indicate systematic uncertainties. The colored boxes at $\RAA=1$ represent fractional uncertainty in $\langle\TAA\rangle$ and $pp$ luminosity in this measurement and $pp$ luminosity in previous analysis~\cite{HION-2017-10}, which both affect the overall normalization. }
\label{fig:pt}
\end{figure}

A systematic comparison of jet suppression in terms of jet \RAA(\pt) for different jet definitions is provided in Figure~\ref{fig:pt}. The jet \RAA\ is measured for:
$R=1.0$ re-clustered jets with single sub-jet, $R=1.0$ re-clustered jets with multiple sub-jets, $R=1.0$ re-clustered inclusive jets, $R=0.2$ jets, and $R=0.4$ jets~\cite{HION-2017-10}.
 
Production of $R=1.0$ re-clustered inclusive jets is suppressed more than the production of $R=0.2$ or $R=0.4$ jets. Various theoretical calculations of the jet quenching where the jet energy is distributed to soft particles predict less suppression when expanding the jet radius by recovering more lost energy, see e.g. Ref.~\cite{Chien:2015hda}. This energy recovery may however not happen in the case of re-clustered \lR\ jets where energy radiated outside of $R=0.2$ sub-jets is removed. A singular situation when the re-clustered \lR\ jet is completely removed due to all sub-jets being suppressed below the 35~\GeV\ cut is very unlikely given the minimum \pT\ threshold of 158 \GeV\ on the $R=1.0$ jet and relatively small multiplicity of sub-jets. Thus, in general, variations in the jet definition used in this study allow including different fractions of the lost energy and energy from the medium response to the showering process. Comparing new results with theoretical calculations may therefore help to understand possible biases in the quantification of energy loss using traditional, small-radius jets.
The presented quantification of cone-size dependence of \RAA\ may also be used to constrain theoretical uncertainties in the recently analytically calculated cone-size dependent energy loss~\cite{Mehtar-Tani:2021fud}.
 
The $R=1.0$ re-clustered jets with multiple sub-jets show the largest suppression while the $R=1.0$ re-clustered jets with a single sub-jet show the smallest suppression out of all jet definitions. The sizable difference between the \RAA for multiple sub-jet and single sub-jet  configurations provides an important input for understanding the role of color decoherence in the jet quenching.
This sizable difference is not due to missing a contribution from the energy radiated out of the $R=0.2$ cone since this radiated energy is not measured in all configurations of $R=1.0$ jets with sub-jets.
The lack of jet \pT\ dependence in the \RAA\ of $R=1.0$ re-clustered jets with multiple sub-jets might be understood to be the consequence of increasing sub-jet multiplicity with increasing jet \pT\ (see sub-jet multiplicity distributions in Appendix). More sub-jets may imply more sources of radiation which may lead to flattening  of \RAA\ at high-\pt.
Presented results can be also used to obtain ratios of per-jet normalized substructure distributions for $R=1.0$ jets by dividing distributions in Figure~\ref{fig:thetwo} by the inclusive $R=1.0$ jet $\RAA$ in Figure~\ref{fig:pt}.
Finally, it should be noted that any direct comparisons of \RAA\ between different jet definitions should be treated with care as the same particles reconstructed with a different procedure might appear in different jet \pT\ intervals.
 
In conclusion, this Letter provides a measurement of the jet nuclear modification factor for \lR\ jets which is differential in \pt and in transverse and radial substructure observables. Presented observations are qualitatively consistent with the hypothesis that jets with hard internal splittings lose more energy, and provide a new perspective for understanding the role of jet structure in jet suppression in the QGP.
 
\FloatBarrier
 
\section*{Acknowledgements}


We thank CERN for the very successful operation of the LHC, as well as the
support staff from our institutions without whom ATLAS could not be
operated efficiently.
 
We acknowledge the support of
ANPCyT, Argentina;
YerPhI, Armenia;
ARC, Australia;
BMWFW and FWF, Austria;
ANAS, Azerbaijan;
CNPq and FAPESP, Brazil;
NSERC, NRC and CFI, Canada;
CERN;
ANID, Chile;
CAS, MOST and NSFC, China;
Minciencias, Colombia;
MEYS CR, Czech Republic;
DNRF and DNSRC, Denmark;
IN2P3-CNRS and CEA-DRF/IRFU, France;
SRNSFG, Georgia;
BMBF, HGF and MPG, Germany;
GSRI, Greece;
RGC and Hong Kong SAR, China;
ISF and Benoziyo Center, Israel;
INFN, Italy;
MEXT and JSPS, Japan;
CNRST, Morocco;
NWO, Netherlands;
RCN, Norway;
MEiN, Poland;
FCT, Portugal;
MNE/IFA, Romania;
MESTD, Serbia;
MSSR, Slovakia;
ARRS and MIZ\v{S}, Slovenia;
DSI/NRF, South Africa;
MICINN, Spain;
SRC and Wallenberg Foundation, Sweden;
SERI, SNSF and Cantons of Bern and Geneva, Switzerland;
MOST, Taiwan;
TENMAK, T\"urkiye;
STFC, United Kingdom;
DOE and NSF, United States of America.
In addition, individual groups and members have received support from
BCKDF, CANARIE, Compute Canada and CRC, Canada;
PRIMUS 21/SCI/017 and UNCE SCI/013, Czech Republic;
COST, ERC, ERDF, Horizon 2020 and Marie Sk{\l}odowska-Curie Actions, European Union;
Investissements d'Avenir Labex, Investissements d'Avenir Idex and ANR, France;
DFG and AvH Foundation, Germany;
Herakleitos, Thales and Aristeia programmes co-financed by EU-ESF and the Greek NSRF, Greece;
BSF-NSF and MINERVA, Israel;
Norwegian Financial Mechanism 2014-2021, Norway;
NCN and NAWA, Poland;
La Caixa Banking Foundation, CERCA Programme Generalitat de Catalunya and PROMETEO and GenT Programmes Generalitat Valenciana, Spain;
G\"{o}ran Gustafssons Stiftelse, Sweden;
The Royal Society and Leverhulme Trust, United Kingdom.
 
The crucial computing support from all WLCG partners is acknowledged gratefully, in particular from CERN, the ATLAS Tier-1 facilities at TRIUMF (Canada), NDGF (Denmark, Norway, Sweden), CC-IN2P3 (France), KIT/GridKA (Germany), INFN-CNAF (Italy), NL-T1 (Netherlands), PIC (Spain), ASGC (Taiwan), RAL (UK) and BNL (USA), the Tier-2 facilities worldwide and large non-WLCG resource providers. Major contributors of computing resources are listed in Ref.~\cite{ATL-SOFT-PUB-2023-001}.


\clearpage

\clearpage
\part*{Appendix}
 
\begin{figure}[h]
\centering
\includegraphics[width=0.49\textwidth]{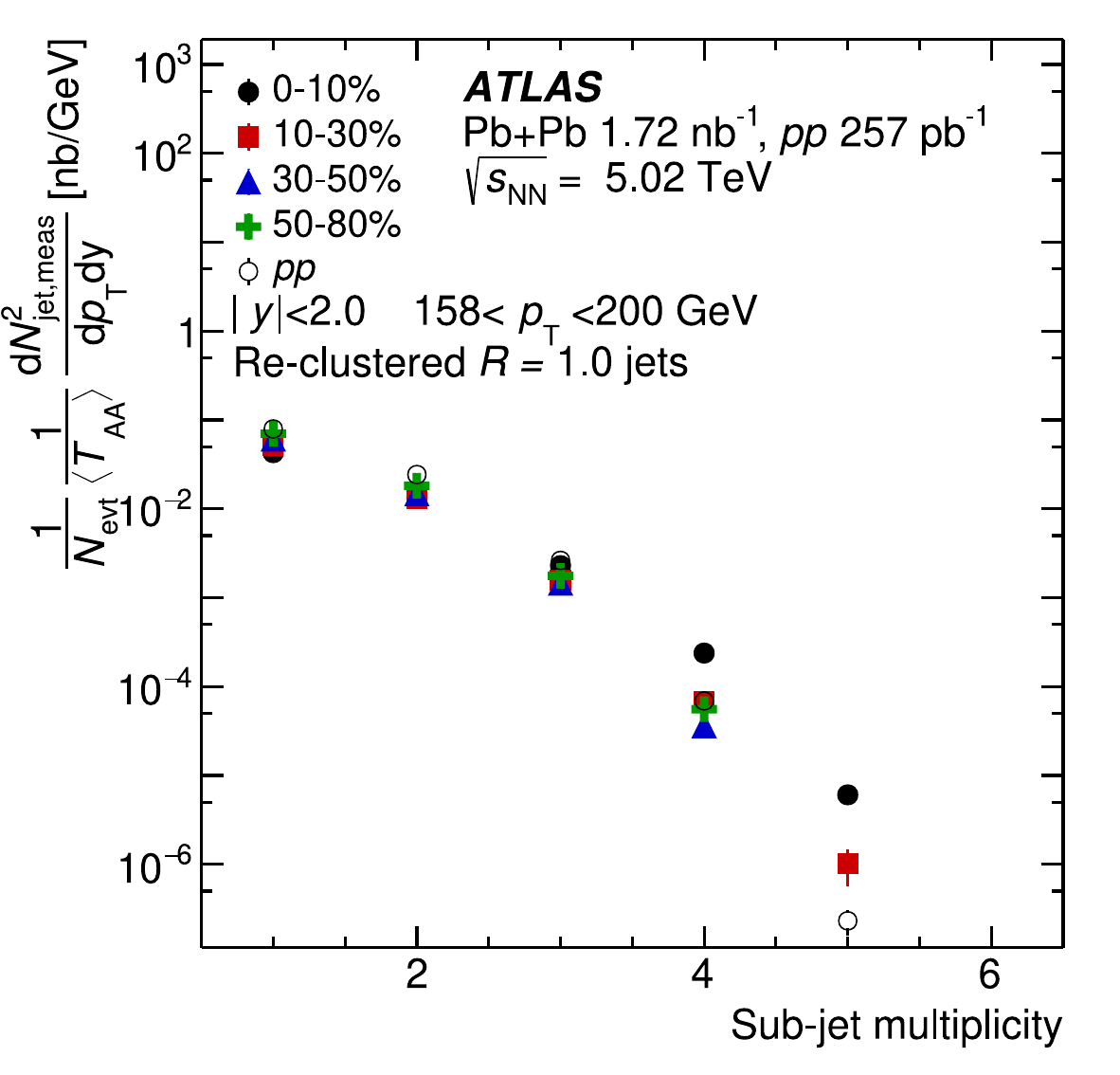}
\includegraphics[width=0.49\textwidth]{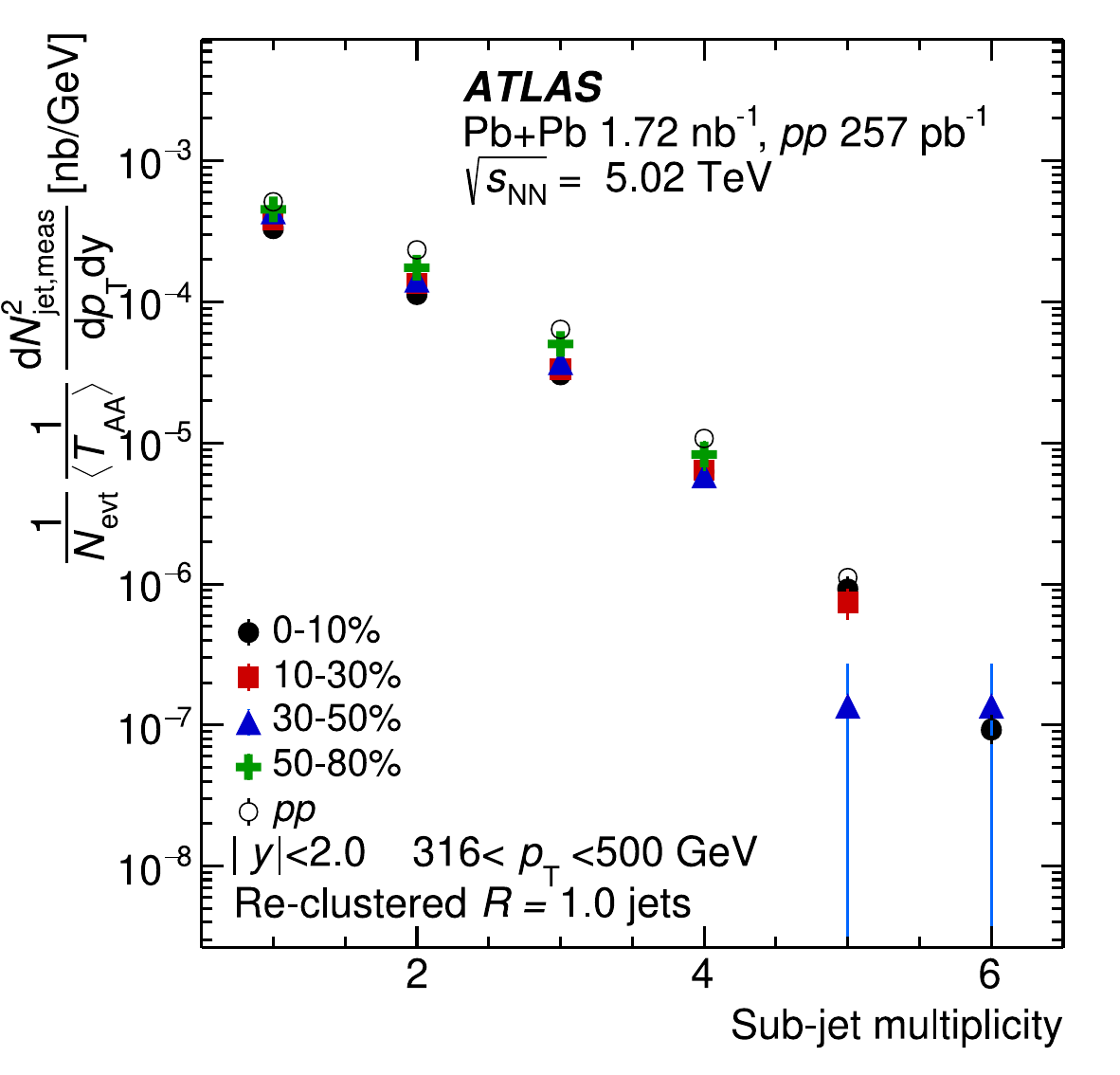}
\caption{
Measured uncorrected distributions of the number of sub-jets of the \lR\ jets in \pp\ collisions and four centrality intervals of \PbPb\ collisions for jet \pT\ in the range 158--200~\GeV\ (left) and 316--500~\GeV\ (right).
}
\label{fig:subjetmultiplicity}
\end{figure}

The measured uncorrected distribution of the yield of jets evaluated as a function of a number of sub-jets with \pT\ above 35 GeV is shown in Figure~\ref{fig:subjetmultiplicity}. The yield is shown for \lR\ jets measured in \pp\ collisions and in four centrality intervals of \PbPb\ collisions. The fraction of reconstructed \lR\ jets with a single sub-jet in \pp\ collisions is 75\% in the \pt\ interval of $158-200$~\GeV and 62\% in the \pt\ interval of $316-500$~\GeV. In 0--10\% central \PbPb\ collisions, the fractions are 73\% and 68\%
in the \pt\ interval of $158-200$~\GeV and $316-500$~\GeV, respectively.
In 50--80\% peripheral collisions, the fractions are 77\% and 65\%, in the \pt\ interval of $158-200$~\GeV and $316-500$~\GeV, respectively. The measured distributions in \PbPb\ collisions include a contribution from jets that originate from different hard partonic interactions in the same \PbPb\ collision and from the spurious fluctuations of UE. These two contributions increase with the increasing centrality of the collision.
In general, the observed increase of yield of \lR\ jets with multiple sub-jets with increasing jet \pT\ is expected both in \pp\ and \PbPb\ collisions, since it was previously measured that the yield of neighboring jets that accompany a given higher-\pT\ jet increases with increasing jet \pT~\cite{D0:2012xif,HION-2013-10}.

\clearpage

\printbibliography

@article{CDF:1994zkl,
    author = "Abe, F. and others",
    collaboration = "CDF",
    title = "{Evidence for color coherence in $p\bar{p}$ collisions at $\sqrt{s} = 1.8$ TeV}",
    reportNumber = "FERMILAB-PUB-94-072-E, CDF-ANAL-JET-PUBLIC-2413",
    doi = "10.1103/PhysRevD.50.5562",
    journal = "Phys. Rev. D",
    volume = "50",
    pages = "5562--5579",
    year = "1994"
}

@article{ALICE:2022vsz,
    author = "{ALICE Collaboration}",
    title = "{Measurement of inclusive and leading subjet fragmentation in pp and Pb-Pb collisions at $\sqrt{s_{NN}}$ = 5.02 TeV}",
    eprint = "2204.10270",
    archivePrefix = "arXiv",
    primaryClass = "nucl-ex",
    reportNumber = "CERN-EP-2022-060",
    month = "4",
    year = "2022"
}

@article{Nachman_2015,
	doi = {10.1007/jhep02(2015)075}, 
	url = {https://doi.org/10.1007%2Fjhep02%282015%29075},
	year = 2015,
	month = {feb},
	publisher = {Springer Science and Business Media {LLC}},
	volume = {2015},
	number = {2},
	author = {Benjamin Nachman and Pascal Nef and Ariel Schwartzman and Maximilian Swiatlowski and Chaowaroj Wanotayaroj}, 
	title = {Jets from jets: re-clustering as a tool for large radius jet reconstruction and grooming at the {LHC}},
	journal = {Journal of High Energy Physics}
}

@Article{HION-2021-09,
    author         = "{ATLAS Collaboration}",
    title          = "{Measurement of substructure-dependent jet suppression in Pb+Pb collisions at 5.02 \text{TeV} with the ATLAS detector}",
    year           = "2022",
    reportNumber   = "CERN-EP-2022-182",
    eprint         = "2211.11470",
    archivePrefix  = "arXiv",
    primaryClass   = "nucl-ex",
}

@article{Cunqueiro:2021wls,
    author = "Cunqueiro, Leticia and Sickles, Anne M.",
    title = "{Studying the QGP with Jets at the LHC and RHIC}",
    eprint = "2110.14490",
    archivePrefix = "arXiv",
    primaryClass = "nucl-ex",
    doi = "10.1016/j.ppnp.2022.103940",
    journal = "Prog. Part. Nucl. Phys.",
    volume = "124",
    pages = "103940",
    year = "2022"
}

@article{D0:2012xif,
    author = "{D$\emptyset$ Collaboration}",
    collaboration = "D0",
    title = "{Measurement of angular correlations of jets at $\sqrt{s}=1.96$ TeV and determination of the strong coupling at high momentum transfers}",
    eprint = "1207.4957",
    archivePrefix = "arXiv",
    primaryClass = "hep-ex",
    reportNumber = "FERMILAB-PUB-12-397-E",
    doi = "10.1016/j.physletb.2012.10.003",
    journal = "Phys. Lett. B",
    volume = "718",
    pages = "56--63",
    year = "2012"
}

@article{ALICE:2021obz,
    author = "{ALICE Collaboration}",
    title = "{Measurement of the groomed jet radius and momentum splitting fraction in pp and Pb$-$Pb collisions at $\sqrt{s_{NN}} = 5.02$ TeV}",
    eprint = "2107.12984",
    archivePrefix = "arXiv",
    primaryClass = "nucl-ex",
    reportNumber = "CERN-EP-2021-151",
    doi = "10.1103/PhysRevLett.128.102001",
    journal = "Phys. Rev. Lett.",
    volume = "128",
    number = "10",
    pages = "102001",
    year = "2022"
}

@article{Heinz:2013th,
    author = "Heinz, Ulrich and Snellings, Raimond",
    title = "{Collective Flow and Viscosity in Relativistic Heavy-Ion Collisions}",
    eprint = "1301.2826",
    archivePrefix = "arXiv",
    primaryClass = "nucl-th",
    doi = "10.1146/annurev-nucl-102212-170540",
    journal = "Ann. Rev. Nucl. Part. Sci.",
    volume = "63",
    pages = "123--151",
    year = "2013"
}

@article{Busza:2018rrf,
    author = "Busza, Wit and Rajagopal, Krishna and van der Schee, Wilke",
    title = "{Heavy Ion Collisions: The Big Picture, and the Big Questions}",
    eprint = "1802.04801",
    archivePrefix = "arXiv",
    primaryClass = "hep-ph",
    reportNumber = "MIT-CTP-4892, MIT-CTP/4892",
    doi = "10.1146/annurev-nucl-101917-020852",
    journal = "Ann. Rev. Nucl. Part. Sci.",
    volume = "68",
    pages = "339--376",
    year = "2018"
}

@article{CasalderreySolana:2012ef,
      author         = "Casalderrey-Solana, Jorge and Mehtar-Tani, Yacine and
                        Salgado, Carlos A. and Tywoniuk, Konrad",
      title          = "{New picture of jet quenching dictated by color
                        coherence}",
      journal        = "Phys. Lett. B",
      volume         = "725",
      year           = "2013",
      pages          = "357-360",
      doi            = "10.1016/j.physletb.2013.07.046",
      eprint         = "1210.7765",
      archivePrefix  = "arXiv",
      primaryClass   = "hep-ph",
      reportNumber   = "CERN-PH-TH-2012-291, LU-TP-12-37",
      SLACcitation   = "%%CITATION = ARXIV:1210.7765;%%"
}

@article{DAgostini:1994zf,
     author         = "D'Agostini, G.",
     title          = "{A multidimensional unfolding method based on Bayes'
                        theorem}",
      journal        = "Nucl.~Instrum.~Meth.~A",
      volume         = "362",
      pages          = "487-498",
      doi            = "10.1016/0168-9002(95)00274-X",
      year           = "1995",
      reportNumber   = "DESY-94-099",
      SLACcitation   = "%%CITATION = NUIMA,A362,487;%%",
}

@article{Qin:2015srf,
      author         = "Qin, Guang-You and Wang, Xin-Nian",
      title          = "{Jet quenching in high-energy heavy-ion collisions}",
      journal        = "Int. J. Mod. Phys.~E",
      volume         = "24",
      year           = "2015",
      number         = "11",
      pages          = "1530014",
      doi            = "10.1142/S0218301315300143",
      eprint         = "1511.00790",
      archivePrefix  = "arXiv",
      primaryClass   = "hep-ph",
      SLACcitation   = "%%CITATION = ARXIV:1511.00790;%%"
}

@article{Miller:2007ri,
      author         = "Miller, Michael L. and Reygers, Klaus and Sanders,
                        Stephen J. and Steinberg, Peter",
      title          = "{Glauber Modeling in High-Energy Nuclear Collisions}",
      journal        = "Ann.Rev.Nucl.Part.Sci.",
      volume         = "57",
      pages          = "205-243",
      doi            = "10.1146/annurev.nucl.57.090506.123020",
      year           = "2007",
      eprint         = "nucl-ex/0701025",
      archivePrefix  = "arXiv",
      primaryClass   = "nucl-ex",
      SLACcitation   = "%%CITATION = NUCL-EX/0701025;%%",
}

@article{Chien:2015hda,
      author         = "Chien, Yang-Ting and Vitev, Ivan",
      title          = "{Towards the understanding of jet shapes and cross
                        sections in heavy ion collisions using soft-collinear
                        effective theory}",
      journal        = "JHEP",
      volume         = "05",
      year           = "2016",
      pages          = "023",
      doi            = "10.1007/JHEP05(2016)023",
      eprint         = "1509.07257",
      archivePrefix  = "arXiv",
      primaryClass   = "hep-ph",
      SLACcitation   = "%%CITATION = ARXIV:1509.07257;%%"
}

@article{Casalderrey-Solana:2017mjg,
      author         = "Casalderrey-Solana, Jorge and Mehtar-Tani, Yacine and
                        Salgado, Carlos A. and Tywoniuk, Konrad",
      title          = "{Probing jet decoherence in heavy ion collisions}",
      booktitle      = "{Proceedings, 26th International Conference on
                        Ultra-relativistic Nucleus-Nucleus Collisions (Quark
                        Matter 2017): Chicago, Illinois, USA, February 5-11,
                        2017}",
      journal        = "Nucl. Phys. A",
      volume         = "967",
      year           = "2017",
      pages          = "564-567",
      doi            = "10.1016/j.nuclphysa.2017.07.008",
      SLACcitation   = "%%CITATION = NUPHA,A967,564;%%"
}

@article{MehtarTani:2011tz,
      author         = "Mehtar-Tani, Y. and Salgado, C. A. and Tywoniuk, K.",
      title          = "{Jets in QCD media: From color coherence to decoherence}",
      journal        = "Phys. Lett. B",
      volume         = "707",
      year           = "2012",
      pages          = "156-159",
      doi            = "10.1016/j.physletb.2011.12.042",
      eprint         = "1102.4317",
      archivePrefix  = "arXiv",
      primaryClass   = "hep-ph",
      SLACcitation   = "%%CITATION = ARXIV:1102.4317;%%"
}

@article{Dasgupta:2013ihk,
    author = "Dasgupta, Mrinal and Fregoso, Alessandro and Marzani, Simone and Salam, Gavin P.",
    title = "{Towards an understanding of jet substructure}",
    eprint = "1307.0007",
    archivePrefix = "arXiv",
    primaryClass = "hep-ph",
    reportNumber = "CERN-PH-TH-2013-145, DCPT-13-86, IPPP-13-43, LPN13-036, MAN-HEP-2013-12",
    doi = "10.1007/JHEP09(2013)029",
    journal = "JHEP",
    volume = "09",
    pages = "029",
    year = "2013"
}

@article{Mehtar-Tani:2021fud,
    author = "Mehtar-Tani, Yacine and Pablos, Daniel and Tywoniuk, Konrad",
    title = "{Cone-Size Dependence of Jet Suppression in Heavy-Ion Collisions}",
    eprint = "2101.01742",
    archivePrefix = "arXiv",
    primaryClass = "hep-ph",
    doi = "10.1103/PhysRevLett.127.252301",
    journal = "Phys. Rev. Lett.",
    volume = "127",
    number = "25",
    pages = "252301",
    year = "2021"
}

@article{Larkoski:2014wba,
    author = "Larkoski, Andrew J. and Marzani, Simone and Soyez, Gregory and Thaler, Jesse",
    title = "{Soft drop}",
    eprint = "1402.2657",
    archivePrefix = "arXiv",
    primaryClass = "hep-ph",
    reportNumber = "MIT-CTP-4531, DCPT-14-24, IPPP-14-12",
    doi = "10.1007/JHEP05(2014)146",
    journal = "JHEP",
    volume = "05",
    pages = "146",
    year = "2014"
}

@article{Tachibana:2019hrn,
    author = "Tachibana, Yasuki",
    editor = "Antinori, Federico and Dainese, Andrea and Giubellino, Paolo and Greco, Vincenzo and Lombardo, Maria Paola and Scomparin, Enrico",
    title = "{Medium response to jet-induced excitation: theory overview}",
    doi = "10.1016/j.nuclphysa.2018.10.074",
    journal = "Nucl. Phys. A",
    volume = "982",
    pages = "156--162",
    year = "2019"
}

@article{Blaizot:2012fh,
    author = "Blaizot, Jean-Paul and Dominguez, Fabio and Iancu, Edmond and Mehtar-Tani, Yacine",
    title = "{Medium-induced gluon branching}",
    eprint = "1209.4585",
    archivePrefix = "arXiv",
    primaryClass = "hep-ph",
    doi = "10.1007/JHEP01(2013)143",
    journal = "JHEP",
    volume = "01",
    pages = "143",
    year = "2013"
}

@book{Marzani:2019hun,
    author = "Marzani, Simone and Soyez, Gregory and Spannowsky, Michael",
    title = "{Looking inside jets: an introduction to jet substructure and boosted-object phenomenology}",
    eprint = "1901.10342",
    archivePrefix = "arXiv",
    primaryClass = "hep-ph",
    doi = "10.1007/978-3-030-15709-8",
    publisher = "Springer",
    volume = "958",
    year = "2019"
}

@article{Butterworth:2008iy,
    author = "Butterworth, Jonathan M. and Davison, Adam R. and Rubin, Mathieu and Salam, Gavin P.",
    title = "{Jet Substructure as a New Higgs-Search Channel at the Large Hadron Collider}",
    eprint = "0802.2470",
    archivePrefix = "arXiv",
    primaryClass = "hep-ph",
    doi = "10.1103/PhysRevLett.100.242001",
    journal = "Phys. Rev. Lett.",
    volume = "100",
    pages = "242001",
    year = "2008"
}

@article{Ringer:2019rfk,
    author = "Ringer, Felix and Xiao, Bo-Wen and Yuan, Feng",
    title = "{Can we observe jet $P_T$-broadening in heavy-ion collisions at the LHC?}",
    eprint = "1907.12541",
    archivePrefix = "arXiv",
    primaryClass = "hep-ph",
    doi = "10.1016/j.physletb.2020.135634",
    journal = "Phys. Lett. B",
    volume = "808",
    pages = "135634",
    year = "2020"
}

@article{Mehtar-Tani:2010ebp,
    author = "Mehtar-Tani, Yacine and Salgado, Carlos A. and Tywoniuk, Konrad",
    title = "{Antiangular Ordering of Gluon Radiation in QCD Media}",
    eprint = "1009.2965",
    archivePrefix = "arXiv",
    primaryClass = "hep-ph",
    doi = "10.1103/PhysRevLett.106.122002",
    journal = "Phys. Rev. Lett.",
    volume = "106",
    pages = "122002",
    year = "2011"
}

@article{ALICE:2021vrw,
    author = "{ALICE Collaboration}",
    collaboration = "ALICE",
    title = "{First measurements of $N$-subjettiness in central Pb$-$Pb collisions at ${\sqrt{s_{\text{NN}}}} = 2.76$ TeV}",
    eprint = "2105.04936",
    archivePrefix = "arXiv",
    primaryClass = "nucl-ex",
    reportNumber = "CERN-EP-2021-082",
    doi = "10.1007/JHEP10(2021)003",
    journal = "JHEP",
    volume = "10",
    pages = "003",
    year = "2021"
}

@article{ALICE:2019ykw,
    author = "{ALICE Collaboration}",
    collaboration = "ALICE",
    title = "{Exploration of jet substructure using iterative declustering in pp and Pb\textendash{}Pb collisions at LHC energies}",
    eprint = "1905.02512",
    archivePrefix = "arXiv",
    primaryClass = "nucl-ex",
    reportNumber = "CERN-EP-2019-087",
    doi = "10.1016/j.physletb.2020.135227",
    journal = "Phys. Lett. B",
    volume = "802",
    pages = "135227",
    year = "2020"
}

@article{Gelis:2021zmx,
    author = "Gelis, Francois",
    title = "{Some aspects of the theory of heavy ion collisions}",
    eprint = "2102.07604",
    archivePrefix = "arXiv",
    primaryClass = "hep-ph",
    doi = "10.1088/1361-6633/abec2e",
    journal = "Rept. Prog. Phys.",
    volume = "84",
    number = "5",
    pages = "056301",
    year = "2021"
}

@article{Andrews:2018jcm,
    author = "Andrews, Harry Arthur and others",
    title = "{Novel tools and observables for jet physics in heavy-ion collisions}",
    eprint = "1808.03689",
    archivePrefix = "arXiv",
    primaryClass = "hep-ph",
    reportNumber = "CERN-TH-2018-186, LU-TP 18-14, IFJPAN-IV-2018-8, MCNET-18-19",
    doi = "10.1088/1361-6471/ab7cbc",
    journal = "J. Phys. G",
    volume = "47",
    number = "6",
    pages = "065102",
    year = "2020"
}

@article{Blaizot:2015lma,
      author         = "Blaizot, Jean-Paul and Mehtar-Tani, Yacine",
      title          = "{Jet structure in heavy ion collisions}",
      journal        = "Int. J. Mod. Phys. E",
      volume         = "24",
      year           = "2015",
      number         = "11",
      pages          = "1530012",
      doi            = "10.1142/S021830131530012X",
      eprint         = "1503.05958",
      archivePrefix  = "arXiv",
      primaryClass   = "hep-ph",
      reportNumber   = "INT-PUB-15-006",
      SLACcitation   = "%%CITATION = ARXIV:1503.05958;%%"
}

@article{Loizides:2014vua,
      author         = "Loizides, C. and Nagle, J. and Steinberg, P.",
      title          = "{Improved version of the PHOBOS Glauber Monte Carlo}",
      journal        = "SoftwareX",
      volume         = "1-2",
      year           = "2015",
      pages          = "13-18",
      doi            = "10.1016/j.softx.2015.05.001",
      eprint         = "1408.2549",
      archivePrefix  = "arXiv",
      primaryClass   = "nucl-ex",
      SLACcitation   = "%%CITATION = ARXIV:1408.2549;%%"
}

@article{Caucal:2018dla,
    author = "Caucal, P. and Iancu, E. and Mueller, A. H. and Soyez, G.",
    title = "{Vacuumlike Jet Fragmentation in a Dense QCD Medium}",
    eprint = "1801.09703",
    archivePrefix = "arXiv",
    primaryClass = "hep-ph",
    doi = "10.1103/PhysRevLett.120.232001",
    journal = "Phys. Rev. Lett.",
    volume = "120",
    pages = "232001",
    year = "2018"
}

@article{Mehtar-Tani:2017web,
    author = "Mehtar-Tani, Yacine and Tywoniuk, Konrad",
    title = "{Sudakov suppression of jets in QCD media}",
    eprint = "1707.07361",
    archivePrefix = "arXiv",
    primaryClass = "hep-ph",
    reportNumber = "INT-PUB-17-025, CERN-TH-2017-161",
    doi = "10.1103/PhysRevD.98.051501",
    journal = "Phys. Rev. D",
    volume = "98",
    number = "5",
    pages = "051501",
    year = "2018"
}

@article{Gyulassy:2000er,	c="GLV2",
      author         = "Gyulassy, M. and Levai, P. and Vitev, I.",
      title          = "{Reaction operator approach to nonAbelian energy loss}",
      journal        = "Nucl. Phys.",
      volume         = "B594",
      year           = "2001",
      pages          = "371-419",
      doi            = "10.1016/S0550-3213(00)00652-0",
      eprint         = "nucl-th/0006010",
      archivePrefix  = "arXiv",
      primaryClass   = "nucl-th",
      reportNumber   = "CU-TP-979",
      SLACcitation   = "%%CITATION = NUCL-TH/0006010;%%"
}

@article{Baier:1996kr,
      author         = "Baier, R. and Dokshitzer, Yuri L. and Mueller, Alfred H.
                        and Peigne, S. and Schiff, D.",
      title          = "{Radiative energy loss of high-energy quarks and gluons
                        in a finite volume quark - gluon plasma}",
      journal        = "Nucl. Phys.",
      volume         = "B483",
      year           = "1997",
      pages          = "291-320",
      doi            = "10.1016/S0550-3213(96)00553-6",
      eprint         = "hep-ph/9607355",
      archivePrefix  = "arXiv",
      primaryClass   = "hep-ph",
      reportNumber   = "BI-TP-96-21, CUTP-759, LPTHE-ORSAY-96-34",
      SLACcitation   = "%%CITATION = HEP-PH/9607355;%%"
}

@article{lhcbcollaboration2022multidifferential,
      title={Multidifferential study of identified charged hadron distributions in $Z$-tagged jets in proton-proton collisions at $\sqrt{s}=$13 TeV}, 
      author="{LHCb Collaboration}",
      year={2022},
      eprint={2208.11691},
      archivePrefix={arXiv},
      primaryClass={hep-ex}
}

@article{Acharya:2020,
      author         = "{ALICE Collaboration}",
      firstauthor    = "Abelev, B. and others",
      title          = "{Measurements of inclusive jet spectra in pp and central Pb-Pb collisions at $\sqrt{s_{NN}}=5.02$ TeV}",
      doi = {10.1103/physrevc.101.034911},  
      year = "2020",
      volume = "101",
      number = "3",
      journal = "Phys. Rev. C",
      eprint         = "1909.09718",
      archivePrefix  = "arXiv",
      primaryClass   = "nucl-ex",
      reportNumber   = "CERN-EP-2019-200",
}

@article{STAR:2020xiv,
    author = "Adam, Jaroslav and others",
    collaboration = "STAR",
    title = "{Measurement of inclusive charged-particle jet production in Au + Au collisions at $\sqrt{s_{NN}}=$200 GeV}",
    eprint = "2006.00582",
    archivePrefix = "arXiv",
    primaryClass = "nucl-ex",
    doi = "10.1103/PhysRevC.102.054913",
    journal = "Phys. Rev. C",
    volume = "102",
    number = "5",
    pages = "054913",
    year = "2020"
}

@article{Catani:1993hr,
	comment	= "kt",
      author         = "Catani, S. and Dokshitzer, Yuri L. and Seymour, M. H. and
                        Webber, B. R.",
      title          = "{Longitudinally-invariant $k_{\perp}$-clustering algorithms for hadron-hadron collisions}",
      journal        = "Nucl. Phys. B",
      volume         = "406",
      year           = "1993",
      pages          = "187-224",
      doi            = "10.1016/0550-3213(93)90166-M",
      reportNumber   = "CERN-TH-6775-93, LU-TP-93-2",
      SLACcitation   = "%%CITATION = NUPHA,B406,187;%%"
}

@Article{Cacciari:2008gp,
     author    = "Cacciari, Matteo and Salam, Gavin P. and Soyez, Gregory",
     title     = "{The anti-\(k_{t}\) jet clustering algorithm}",
     journal   = "JHEP",
     volume    = "04",
     year      = "2008",
     pages     = "063",
     eprint    = "0802.1189",
     archivePrefix = "arXiv",
     primaryClass  =  "hep-ph",
     doi       = "10.1088/1126-6708/2008/04/063",
     SLACcitation  = "%%CITATION = 0802.1189;%%"
}

@Article{Fastjet,
      author         = "Cacciari, Matteo and Salam, Gavin P. and Soyez, Gregory",
      title          = "{FastJet user manual}",
      journal        = "Eur. Phys. J. C",
      volume         = "72",
      year           = "2012",
      pages          = "1896",
      doi            = "10.1140/epjc/s10052-012-1896-2",
      eprint         = "1111.6097",
      archivePrefix  = "arXiv",
      primaryClass   = "hep-ph",
      reportNumber   = "CERN-PH-TH-2011-297",
      SLACcitation   = "%%CITATION = ARXIV:1111.6097;%%"
}

@article{Martin:2009iq,
      author         = "Martin, A. D. and Stirling, W. J. and Thorne, R. S. and Watt, G.",
      title          = "{Parton distributions for the LHC}",
      journal        = "Eur. Phys. J. C",
      volume         = "63",
      year           = "2009",
      pages          = "189",
      doi            = "10.1140/epjc/s10052-009-1072-5",
      eprint         = "0901.0002",
      archivePrefix  = "arXiv",
      primaryClass   = "hep-ph",
      reportNumber   = "IPPP-08-95, DCPT-08-190, CAVENDISH-HEP-08-16",
      SLACcitation   = "%%CITATION = ARXIV:0901.0002;%%"
}

@article{Ball:2012cx,
      author         = "{NNPDF Collaboration} and Ball, Richard D. and others",
      title          = "{Parton distributions with LHC data}",
      journal        = "Nucl. Phys. B",
      volume         = "867",
      year           = "2013",
      pages          = "244",
      doi            = "10.1016/j.nuclphysb.2012.10.003",
      eprint         = "1207.1303",
      archivePrefix  = "arXiv",
      primaryClass   = "hep-ph",
      reportNumber   = "EDINBURGH-2012-08, IFUM-FT-997, FR-PHENO-2012-014,
                        RWTH-TTK-12-25, CERN-PH-TH-2012-037, SFB-CPP-12-47\,
                        --CERN-PH-TH-2012-037",
      SLACcitation   = "%%CITATION = ARXIV:1207.1303;%%"
}

@article{Sjostrand:2014zea,
      author         = "Sj{\"o}strand, Torbj{\"o}rn and Ask, Stefan and Christiansen,
                        Jesper R. and Corke, Richard and Desai, Nishita and Ilten,
                        Philip and Mrenna, Stephen and Prestel, Stefan and
                        Rasmussen, Christine O. and Skands, Peter Z.",
      title          = "{An introduction to PYTHIA 8.2}",
      journal        = "Comput. Phys. Commun.",
      volume         = "191",
      year           = "2015",
      pages          = "159",
      doi            = "10.1016/j.cpc.2015.01.024",
      eprint         = "1410.3012",
      archivePrefix  = "arXiv",
      primaryClass   = "hep-ph",
      reportNumber   = "LU-TP-14-36, MCNET-14-22, CERN-PH-TH-2014-190,
                        FERMILAB-PUB-14-316-CD, DESY-14-178, SLAC-PUB-16122,
                        --FERMILAB-PUB-14-316-CD",
      SLACcitation   = "%%CITATION = ARXIV:1410.3012;%%"
}

@article{Agostinelli:2002hh,
      author         = "Agostinelli, S. and others",
      title          = "{\textsc{Geant4} -- a simulation toolkit}",
      journal        = "Nucl. Instrum. Meth. A",
      volume         = "506",
      year           = "2003",
      pages          = "250",
      doi            = "10.1016/S0168-9002(03)01368-8",
      reportNumber   = "SLAC-PUB-9350, FERMILAB-PUB-03-339",
      SLACcitation   = "%%CITATION = NUIMA,A506,250;%%"
}

@Inproceedings{Adye:2011gm,
    author = "Adye, Tim",
    title = "{Unfolding algorithms and tests using RooUnfold}",
    booktitle = "{Proceedings, 2011 Workshop on Statistical Issues Related to Discovery Claims in Search Experiments and Unfolding (PHYSTAT 2011)}",
    venue = "CERN, Geneva, Switzerland",
    eventdate = "2011-01-17/2011-01-20",
    pages = "313-318",
    eprint = "1105.1160",
    archivePrefix = "arXiv",
    primaryClass = "physics.data-an",
    doi = "10.5170/CERN-2011-006.313",
}

@Article{PERF-2007-01,
    author         = "{ATLAS Collaboration}",
    title          = "{The ATLAS Experiment at the CERN Large Hadron Collider}",
    journal        = "JINST",
    volume         = "3",
    year           = "2008",
    pages          = "S08003",
    doi            = "10.1088/1748-0221/3/08/S08003",
    primaryClass   = "hep-ex",
}

@Article{SOFT-2010-01,
    author         = "{ATLAS Collaboration}",
    title          = "{The ATLAS Simulation Infrastructure}",
    journal        = "Eur. Phys. J. C",
    volume         = "70",
    year           = "2010",
    pages          = "823",
    doi            = "10.1140/epjc/s10052-010-1429-9",
    eprint         = "1005.4568",
    archivePrefix  = "arXiv",
    primaryClass   = "physics.ins-det",
}

@Article{PERF-2011-04,
    author         = "{ATLAS Collaboration}",
    title          = "{Jet energy resolution in proton--proton collisions at \(\sqrt{s} = 7\,\text{TeV}\) recorded in 2010 with the ATLAS detector}",
    journal        = "Eur. Phys. J. C",
    volume         = "73",
    year           = "2013",
    pages          = "2306",
    doi            = "10.1140/epjc/s10052-013-2306-0",
    reportNumber   = "CERN-PH-EP-2012-191",
    eprint         = "1210.6210",
    archivePrefix  = "arXiv",
    primaryClass   = "hep-ex",
}

@Article{HION-2013-06,
    author         = "{ATLAS Collaboration}",
    title          = "{Measurements of the Nuclear Modification Factor for Jets in Pb+Pb Collisions at \(\sqrt{s_{\text{NN}}} = 2.76\,\text{TeV}\) with the ATLAS Detector}",
    journal        = "Phys. Rev. Lett.",
    volume         = "114",
    year           = "2015",
    pages          = "072302",
    doi            = "10.1103/PhysRevLett.114.072302",
    reportNumber   = "CERN-PH-EP-2014-172",
    eprint         = "1411.2357",
    archivePrefix  = "arXiv",
    primaryClass   = "hep-ex",
}

@Article{HION-2013-10,
    author         = "{ATLAS Collaboration}",
    title          = "{Measurement of the production of neighbouring jets in lead-lead collisions at \(\sqrt{s_{\text{NN}}} = 2.76\,\text{TeV}\) with the ATLAS detector}",
    journal        = "Phys. Lett. B",
    volume         = "751",
    year           = "2015",
    pages          = "376",
    doi            = "10.1016/j.physletb.2015.10.059",
    reportNumber   = "CERN-PH-EP-2015-143",
    eprint         = "1506.08656",
    archivePrefix  = "arXiv",
    primaryClass   = "hep-ex",
}

@Article{HION-2016-07,
    author         = "{ATLAS Collaboration}",
    title          = "{Prompt and non-prompt \(J/\psi\) and \(\psi(2\text{S})\) suppression at high transverse momentum in \(5.02\,\text{TeV}\) Pb+Pb collisions with the ATLAS experiment}",
    journal        = "Eur. Phys. J. C",
    volume         = "78",
    year           = "2018",
    pages          = "762",
    doi            = "10.1140/epjc/s10052-018-6219-9",
    reportNumber   = "CERN-EP-2018-049",
    eprint         = "1805.04077",
    archivePrefix  = "arXiv",
    primaryClass   = "hep-ex",
}

@Article{PERF-2016-04,
    author         = "{ATLAS Collaboration}",
    title          = "{Jet energy scale measurements and their systematic uncertainties in proton--proton collisions at \(\sqrt{s} = 13\,\text{TeV}\) with the ATLAS detector}",
    journal        = "Phys. Rev. D",
    volume         = "96",
    year           = "2017",
    pages          = "072002",
    doi            = "10.1103/PhysRevD.96.072002",
    reportNumber   = "CERN-EP-2017-038",
    eprint         = "1703.09665",
    archivePrefix  = "arXiv",
    primaryClass   = "hep-ex",
}

@Article{TRIG-2016-01,
    author         = "{ATLAS Collaboration}",
    title          = "{Performance of the ATLAS trigger system in 2015}",
    journal        = "Eur. Phys. J. C",
    volume         = "77",
    year           = "2017",
    pages          = "317",
    doi            = "10.1140/epjc/s10052-017-4852-3",
    reportNumber   = "CERN-EP-2016-241",
    eprint         = "1611.09661",
    archivePrefix  = "arXiv",
    primaryClass   = "hep-ex",
}

@Article{HION-2017-04,
    author         = "{ATLAS Collaboration}",
    title          = "{Measurement of jet fragmentation in Pb+Pb and \(pp\) collisions at \(\sqrt{s_{\text{NN}}} = 5.02\,\text{TeV}\) with the ATLAS detector}",
    journal        = "Phys. Rev. C",
    volume         = "98",
    year           = "2018",
    pages          = "024908",
    doi            = "10.1103/PhysRevC.98.024908",
    reportNumber   = "CERN-EP-2018-096",
    eprint         = "1805.05424",
    archivePrefix  = "arXiv",
    primaryClass   = "hep-ex",
}

@Article{HION-2017-10,
    author         = "{ATLAS Collaboration}",
    title          = "{Measurement of the nuclear modification factor for inclusive jets in Pb+Pb collisions at \(\sqrt{s_{\text{NN}}} = 5.02\,\text{TeV}\) with the ATLAS detector}",
    journal        = "Phys. Lett. B",
    volume         = "790",
    year           = "2019",
    pages          = "108",
    doi            = "10.1016/j.physletb.2018.10.076",
    reportNumber   = "CERN-EP-2018-105",
    eprint         = "1805.05635",
    archivePrefix  = "arXiv",
    primaryClass   = "hep-ex",
}

@Article{HION-2018-03,
    author         = "{ATLAS Collaboration}",
    title          = "{Measurement of angular and momentum distributions of charged particles within and around jets in Pb+Pb and \(pp\) collisions at \(\sqrt{s_{\text{NN}}} = 5.02\,\text{TeV}\) with the ATLAS detector}",
    journal        = "Phys. Rev. C",
    volume         = "100",
    year           = "2019",
    pages          = "064901",
    doi            = "10.1103/PhysRevC.100.064901",
    reportNumber   = "CERN-EP-2019-158",
    eprint         = "1908.05264",
    archivePrefix  = "arXiv",
    primaryClass   = "hep-ex",
    related        = "HION-2018-03-err",
    relatedstring  = "Erratum:",
}

@Article{HION-2018-06,
    author         = "{ATLAS Collaboration}",
    title          = "{Measurement of photon-jet transverse momentum correlations in \(5.02\,\text{TeV}\) Pb+Pb and \(pp\) collisions with ATLAS}",
    journal        = "Phys. Lett. B",
    volume         = "789",
    year           = "2019",
    pages          = "167",
    doi            = "10.1016/j.physletb.2018.12.023",
    reportNumber   = "CERN-EP-2018-196",
    eprint         = "1809.07280",
    archivePrefix  = "arXiv",
    primaryClass   = "hep-ex",
}

@Booklet{ATL-SOFT-PUB-2021-001,
    author         = "{ATLAS Collaboration}",
    title          = "{The ATLAS Collaboration Software and Firmware}",
    howpublished   = "{ATL-SOFT-PUB-2021-001}",
    url            = "https://cds.cern.ch/record/2767187",
    year           = "2021",
}

@Booklet{ATL-SOFT-PUB-2023-001,
    author         = "{ATLAS Collaboration}",
    title          = "{ATLAS Computing Acknowledgements}",
    howpublished   = "{ATL-SOFT-PUB-2023-001}",
    url            = "https://cds.cern.ch/record/2869272",
    year           = "2023",
}

@Article{CMS-HIN-12-002,
    author         = "{CMS Collaboration}",
    title          = "{Modification of jet shapes in PbPb collisions at \(\sqrt{s_{\text{NN}}} = 2.76\,\text{TeV}\)}",
    journal        = "Phys. Lett. B",
    volume         = "730",
    year           = "2014",
    pages          = "243",
    doi            = "10.1016/j.physletb.2014.01.042",
    reportNumber   = "CERN-PH-EP-2013-189",
    eprint         = "1310.0878",
    archivePrefix  = "arXiv",
    primaryClass   = "hep-ex",
}

@Article{CMS-SMP-12-010,
    author         = "{CMS Collaboration}",
    title          = "{Probing color coherence effects in \(pp\) collisions at \(\sqrt{s} = 7\,\text{TeV}\)}",
    journal        = "Eur. Phys. J. C",
    volume         = "74",
    year           = "2014",
    pages          = "2901",
    doi            = "10.1140/epjc/s10052-014-2901-8",
    reportNumber   = "CERN-PH-EP-2013-200",
    eprint         = "1311.5815",
    archivePrefix  = "arXiv",
    primaryClass   = "hep-ex",
}

@Article{CMS-HIN-13-005,
    author         = "{CMS Collaboration}",
    title          = "{Measurement of inclusive jet cross-sections in \(pp\) and PbPb collisions at \(\sqrt{s_{\text{NN}}} = 2.76\,\text{TeV}\)}",
    journal        = "Phys. Rev. C",
    volume         = "96",
    year           = "2017",
    pages          = "015202",
    doi            = "10.1103/PhysRevC.96.015202",
    reportNumber   = "CERN-EP-2016-217",
    eprint         = "1609.05383",
    archivePrefix  = "arXiv",
    primaryClass   = "hep-ex",
}

@Article{CMS-HIN-16-006,
    author         = "{CMS Collaboration}",
    title          = "{Measurement of the Splitting Function in \(pp\) and PbPb collisions at \(\sqrt{s_{\text{NN}}} = 5.02\,\text{TeV}\)}",
    journal        = "Phys. Rev. Lett.",
    volume         = "120",
    year           = "2018",
    pages          = "142302",
    doi            = "10.1103/PhysRevLett.120.142302",
    reportNumber   = "CERN-EP-2017-205",
    eprint         = "1708.09429",
    archivePrefix  = "arXiv",
    primaryClass   = "hep-ex",
}

@Article{CMS-HIN-16-024,
    author         = "{CMS Collaboration}",
    title          = "{Measurement of the groomed jet mass in PbPb and \(pp\) collisions at \(\sqrt{s_{\text{NN}}} = 5.02\,\text{TeV}\)}",
    journal        = "JHEP",
    volume         = "10",
    year           = "2018",
    pages          = "161",
    doi            = "10.1007/JHEP10(2018)161",
    reportNumber   = "CERN-EP-2018-097",
    eprint         = "1805.05145",
    archivePrefix  = "arXiv",
    primaryClass   = "hep-ex",
}

@Booklet{ATLAS-CONF-2015-016,
    author         = "{ATLAS Collaboration}",
    title          = "{Jet energy scale and its uncertainty for jets reconstructed using the ATLAS heavy ion jet algorithm}",
    howpublished   = "{ATLAS-CONF-2015-016}",
    url            = "https://cds.cern.ch/record/2008677",
    year           = "2015",
}

@Booklet{ATLAS-CONF-2015-017,
    author         = "{ATLAS Collaboration}",
    title          = "{Data-driven determination of the energy scale and resolution of jets reconstructed in the ATLAS calorimeters using dijet and multijet events at \(\sqrt{s} = 8~\text{TeV}\)}",
    howpublished   = "{ATLAS-CONF-2015-017}",
    url            = "https://cds.cern.ch/record/2008678",
    year           = "2015",
}

@Booklet{ATLAS-CONF-2020-023,
    author         = "{ATLAS Collaboration}",
    title          = "{Luminosity determination for low-pileup datasets at \(\sqrt{s} = 5\) and \(13\,\text{TeV}\) using the ATLAS detector at the LHC}",
    howpublished   = "{ATLAS-CONF-2020-023}",
    url            = "https://cds.cern.ch/record/2725195",
    year           = "2020",
}

@Booklet{ATL-PHYS-PUB-2011-014,
    author         = "{ATLAS Collaboration}",
    title          = "{Further ATLAS tunes of \textsc{Pythia}~6 and Pythia~8}",
    howpublished   = "{ATL-PHYS-PUB-2011-014}",
    url            = "https://cds.cern.ch/record/1400677",
    year           = "2011",
}

@Booklet{ATL-PHYS-PUB-2014-021,
    author         = "{ATLAS Collaboration}",
    title          = "{ATLAS Pythia~8 tunes to \(7~\text{TeV}\) data}",
    howpublished   = "{ATL-PHYS-PUB-2014-021}",
    url            = "https://cds.cern.ch/record/1966419",
    year           = "2014",
}

@Booklet{ATL-DAQ-PUB-2019-001,
    author         = "{ATLAS Collaboration}",
    title          = "{Trigger Menu in 2018}",
    howpublished   = "{ATL-DAQ-PUB-2019-001}",
    url            = "https://cds.cern.ch/record/2693402",
    year           = "2019",
}
 
\clearpage
 
\begin{flushleft}
\hypersetup{urlcolor=black}
{\Large The ATLAS Collaboration}

\bigskip

\AtlasOrcid[0000-0002-6665-4934]{G.~Aad}$^\textrm{\scriptsize 102}$,
\AtlasOrcid[0000-0002-5888-2734]{B.~Abbott}$^\textrm{\scriptsize 120}$,
\AtlasOrcid[0000-0002-1002-1652]{K.~Abeling}$^\textrm{\scriptsize 55}$,
\AtlasOrcid[0000-0002-8496-9294]{S.H.~Abidi}$^\textrm{\scriptsize 29}$,
\AtlasOrcid[0000-0002-9987-2292]{A.~Aboulhorma}$^\textrm{\scriptsize 35e}$,
\AtlasOrcid[0000-0001-5329-6640]{H.~Abramowicz}$^\textrm{\scriptsize 151}$,
\AtlasOrcid[0000-0002-1599-2896]{H.~Abreu}$^\textrm{\scriptsize 150}$,
\AtlasOrcid[0000-0003-0403-3697]{Y.~Abulaiti}$^\textrm{\scriptsize 117}$,
\AtlasOrcid[0000-0003-0762-7204]{A.C.~Abusleme~Hoffman}$^\textrm{\scriptsize 137a}$,
\AtlasOrcid[0000-0002-8588-9157]{B.S.~Acharya}$^\textrm{\scriptsize 69a,69b,p}$
% ...
\bigskip
\\

% ...
$^{29}$Physics Department, Brookhaven National Laboratory, Upton NY; United States of America.\\
% ...
$^{35}$$^{(a)}$Facult\'e des Sciences Ain Chock, R\'eseau Universitaire de Physique des Hautes Energies - Universit\'e Hassan II, Casablanca;$^{(b)}$Facult\'{e} des Sciences, Universit\'{e} Ibn-Tofail, K\'{e}nitra;$^{(c)}$Facult\'e des Sciences Semlalia, Universit\'e Cadi Ayyad, LPHEA-Marrakech;$^{(d)}$LPMR, Facult\'e des Sciences, Universit\'e Mohamed Premier, Oujda;$^{(e)}$Facult\'e des sciences, Universit\'e Mohammed V, Rabat;$^{(f)}$Institute of Applied Physics, Mohammed VI Polytechnic University, Ben Guerir; Morocco.\\
% ...
$^{55}$II. Physikalisches Institut, Georg-August-Universit\"{a}t G\"ottingen, G\"ottingen; Germany.\\
% ...
$^{69}$$^{(a)}$INFN Gruppo Collegato di Udine, Sezione di Trieste, Udine;$^{(b)}$ICTP, Trieste;$^{(c)}$Dipartimento Politecnico di Ingegneria e Architettura, Universit\`a di Udine, Udine; Italy.\\
% ...
$^{102}$CPPM, Aix-Marseille Universit\'e, CNRS/IN2P3, Marseille; France.\\
% ...
$^{117}$Department of Physics, New York University, New York NY; United States of America.\\
% ...
$^{120}$Homer L. Dodge Department of Physics and Astronomy, University of Oklahoma, Norman OK; United States of America.\\
% ...
$^{137}$$^{(a)}$Departamento de F\'isica, Pontificia Universidad Cat\'olica de Chile, Santiago;$^{(b)}$Millennium Institute for Subatomic physics at high energy frontier (SAPHIR), Santiago;$^{(c)}$Instituto de Investigaci\'on Multidisciplinario en Ciencia y Tecnolog\'ia, y Departamento de F\'isica, Universidad de La Serena;$^{(d)}$Universidad Andres Bello, Department of Physics, Santiago;$^{(e)}$Instituto de Alta Investigaci\'on, Universidad de Tarapac\'a, Arica;$^{(f)}$Departamento de F\'isica, Universidad T\'ecnica Federico Santa Mar\'ia, Valpara\'iso; Chile.\\
% ...
$^{150}$Department of Physics, Technion, Israel Institute of Technology, Haifa; Israel.\\
$^{151}$Raymond and Beverly Sackler School of Physics and Astronomy, Tel Aviv University, Tel Aviv; Israel.\\
% ...
$^{a}$ Also Affiliated with an institute covered by a cooperation agreement with CERN.\\
$^{b}$ Also at An-Najah National University, Nablus; Palestine.\\
$^{c}$ Also at Borough of Manhattan Community College, City University of New York, New York NY; United States of America.\\
$^{d}$ Also at Bruno Kessler Foundation, Trento; Italy.\\
$^{e}$ Also at Center for High Energy Physics, Peking University; China.\\
$^{f}$ Also at Center for Interdisciplinary Research and Innovation (CIRI-AUTH), Thessaloniki; Greece.\\
% ...
$^{p}$ Also at Department of Physics, King's College London, London; United Kingdom.\\
% ...

\end{flushleft}


\clearpage

\end{document}